\documentclass[twocolumn]{aastex63}

\received{XXXX}
\revised{XXXX}
\accepted{XXXX}
\submitjournal{ApJ}

\shorttitle{Reliability of a method to map the chromospheric magnetic field azimuth}

\shortauthors{Jur\v{c}\'{a}k et al.}

\begin{document}

\title{Evaluating the reliability of a simple method to map \\the magnetic field azimuth
in the solar chromosphere}

\correspondingauthor{Jan Jur\v{c}\'{a}k}
\email{jan.jurcak@asu.cas.cz}

\author[0000-0002-9220-4515]{Jan Jur\v{c}\'{a}k}
\affiliation{Astronomical Institute of the Czech Academy of Sciences \\
Fri\v{c}ova  298, 25165 Ond\v{r}ejov, Czech Republic}

\author[0000-0002-8292-2636]{Ji\v{r}\'{\i} \v{S}t\v{e}p\'{a}n}
\affiliation{Astronomical Institute of the Czech Academy of Sciences \\
Fri\v{c}ova  298, 25165 Ond\v{r}ejov, Czech Republic}

\author[0000-0001-5131-4139]{Javier Trujillo Bueno}
\affiliation{Instituto de Astrof\'{\i}sica de Canarias \\
E-38205 La Laguna, Tenerife, Spain}
\affiliation{Departamento de Astrof\'{\i}sica, Facultad de F\'{\i}sica, Universidad de La Laguna \\
E-38206 La Laguna, Tenerife, Spain}
\affiliation{Consejo Superior de Investigaciones Cient\'{\i}ficas, Spain}

\begin{abstract}

The Zeeman effect is of limited utility for probing the magnetism of the quiet solar chromosphere. The Hanle effect in some spectral lines is sensitive to such magnetism, but the interpretation of the scattering polarization signals requires taking into account that the chromospheric plasma is highly inhomogeneous and dynamic (i.e., that the magnetic field is not the only cause of symmetry breaking). Here we investigate the reliability of a well-known formula for mapping the azimuth of chromospheric magnetic fields directly from the scattering polarization observed in the \ion{Ca}{2}~8542~\AA\, line, which is typically in the saturation regime of the Hanle effect. To this end, we use the Stokes profiles of the \ion{Ca}{2}~8542~\AA\, line computed with the PORTA radiative transfer code in a three-dimensional (3D) model of the solar chromosphere, degrading them to mimic spectropolarimetric observations for a range of telescope apertures and noise levels. The simulated observations are used to obtain the magnetic field azimuth at each point of the field of view, which we compare with the actual values within the 3D model. We show that, apart from intrinsic ambiguities, the method provides solid results. Their accuracy depends more on the noise level than on the telescope diameter. Large-aperture solar telescopes, like DKIST and EST, are needed to achieve the required noise-to-signal ratios using reasonable exposure times. 

\end{abstract}

\keywords{Sun: chromosphere --- 
techniques: polarimetric --- methods: data analysis}

\section{Introduction}

With the upcoming new generation of solar telescopes, like DKIST \citep[presently in the commissioning phase,][]{Rimmele:2020} and EST \citep[presently in the preparatory phase,][]{Jurcak:2019}, there is an urgent need for suitable methods to infer the magnetic field information from the unprecedented spectropolarimetric data that these telescopes will provide. In particular, reliable diagnostic methods are important for the solar chromosphere, a highly inhomogeneous and dynamic atmospheric region where there are multiple effects that significantly complicate the development of reliable inversion methods. The present paper studies quantitatively the possibility of obtaining realistic maps of the magnetic field azimuth in the solar chromosphere directly from the scattering polarization observed in the \ion{Ca}{2}~8542~\AA\, spectral line. 

To determine the magnetic field in the solar atmosphere, we need to observe 
the intensity ($I$), linear polarization ($Q$ and $U$) and  
circular polarization ($V$) in spectral lines --that is, we need to measure the 
Stokes profiles $I(\lambda)$, $Q(\lambda)$, $U(\lambda)$ and $V(\lambda)$ 
as as function of wavelength \citep[e.g.,][]{Stenflo:1994, Iniesta:2003}.

While the linear polarization of the radiation is most commonly described 
in terms of the Stokes parameters $Q$ and $U$, 
an equivalent description is provided by the total linear polarization, 
$P_{\rm L}=\sqrt{Q^2+U^2}$, and the azimuth $\chi$ of the linear polarization orientation   
with respect to a suitable axis. The simple relation between these two representations is
\begin{eqnarray}
Q&=&P_{\rm L}\cos 2\chi\,, \label{eq:qchi} \\
U&=&P_{\rm L}\sin 2\chi\,,
\end{eqnarray}
which implies that 
\begin{equation}
\chi=\frac 12 \arctan\frac UQ + \chi_0\,,
\label{eq:atan}
\end{equation}
where $\chi_0=0$ if $Q>0$ and $U>0$, $\chi_0=180^\circ$ if $Q>0$ and $U<0$, and $\chi_0=90^\circ$ if $Q<0$
\citep[see Sections 1.1 and 1.8 in][for details]{Egidio:2004}.

\begin{figure*}[t]
    \includegraphics[width=\linewidth]{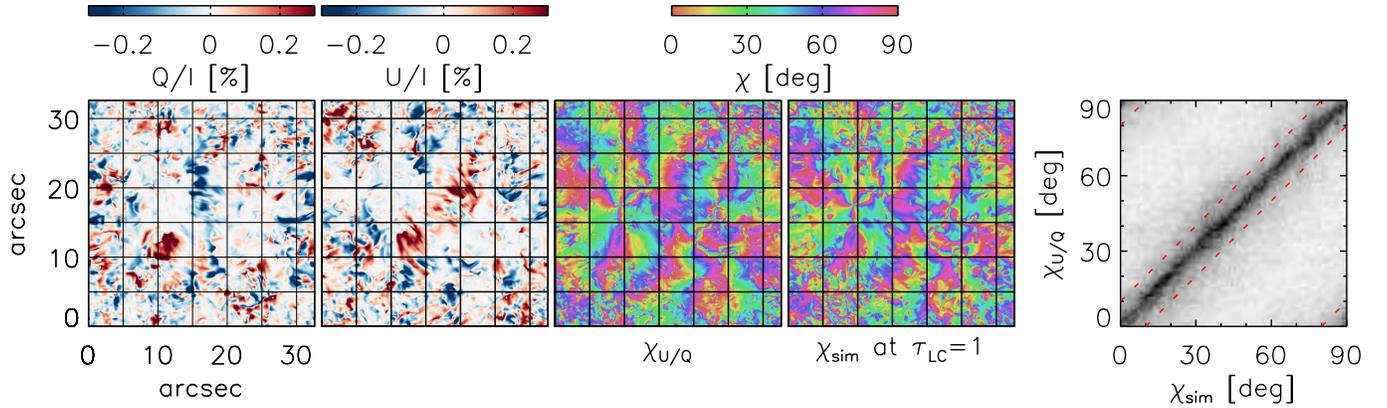}
    \caption{The two left panels show the calculated linear polarization signals integrated over the inner core of the \ion{Ca}{2}~8542~\AA\, line. The next two panels show, respectively, the map of the magnetic field azimuth $\chi_{U/Q}$ determined from the calculated $Q/I$ and $U/I$ signals, and the map of the azimuth $\chi_\mathrm{sim}$ of the magnetic field vector at the corrugated surface within the 3D model where the line-center optical depth is unity in the \ion{Ca}{2}~8542~\AA\, line. The rightmost panel shows the two-dimensional probability density function described in the text, with the red dashed lines indicating the regions where 
    $|\chi_{U/Q} - \chi_\mathrm{sim}| < 10^\circ$.}
   \label{fig1}
\end{figure*}

In solar spectral lines the linear polarization generally results from the scattering of anisotropic radiation and the Zeeman and Hanle effects. When the atomic excitation is dominated by collisional processes or, in other words, when the assumption of local thermodynamic equilibrium (LTE) holds, scattering processes and the Hanle effect do not play any role and the Zeeman effect is the only one that can introduce polarization in the spectral line. Assume that the magnetic field is sufficiently weak, so that the ensuing Zeeman splitting is much smaller than the spectral line width, but sufficiently strong so that the spectral line is in the saturation regime of the Hanle effect (see below), and that its azimuth $\chi_B$ in the observer's reference frame
is constant along the line of sight (LOS). Under such circumstances,
the observed orientation of the linear polarization vector is either parallel or perpendicular to the projection of the magnetic field vector onto the plane of the sky
\citep[e.g., Section~13.5 in][]{Egidio:2004}. Then, it follows from Eqs.~(\ref{eq:qchi}--\ref{eq:atan}) that

\begin{equation}
      \tan 2\chi_B = U(\lambda)/Q(\lambda)\,.
      \label{chi_eq}
\end{equation}
In summary, if a spectral line is in the saturation regime of the Hanle effect and $\chi_B$
is constant along the LOS, Eq.~(\ref{chi_eq}) provides a 
well-known recipe to determine the magnetic field azimuth 
from the orientation of the observed linear polarization.

In chromospheric lines the atomic excitation is typically dominated by radiative transitions. As
a result, scattering processes and the Hanle effect are often the main cause of the spectral line radiation emerging from weak-field regions. Although the scattering polarization in most chromospheric lines is sensitive to magnetic strengths in the gauss range (e.g., the Ca {\sc i} line at 4227~\AA\,), there are a few chromospheric spectral lines that are effectively in the saturation regime of the Hanle effect. In this regime, which occurs when the Zeeman splitting in frequency units is much larger than the inverse lifetime of the relevant atomic levels, the linear polarization that results from scattering processes is sensitive only to the orientation of the magnetic field, but not to its strength \citep[e.g., Section~13.5 in][]{Egidio:2004}. For such spectral lines, like the forbidden lines of the solar corona \citep[e.g.,][]{Judge:2007}, Eq.~(\ref{chi_eq}) holds if the magnetic field azimuth is constant along the LOS.

Interestingly, the well-known \ion{Ca}{2}~8542~\AA\, chromospheric line enters such a regime already for magnetic fields stronger than only a few gauss, because its enigmatic scattering polarization \citep{Stenflo:2000} is dominated by the atomic polarization of the long-lived metastable lower level $4^2{\rm D}_{3/2}$  \citep{manso:2003x,manso:2003,manso:2010}. It is generally believed that such magnetic fields are omnipresent in the quiet solar chromosphere \citep[e.g.,][]{Bianda:1998} and this is also the case in the 
3D model of \citet{Carlsson:2016} we have selected for this investigation.
Other authors have also indicated that such conditions can be expected for the quiet solar chromosphere \cite[e.g., see the left panel of Figure~7 in][]{2015ApJ...801...16C}.

The idea of applying Eq.~(\ref{chi_eq}) to map the azimuth of solar magnetic fields from the forward scattering polarization observed in a spectral line that is close to the saturation regime of the Hanle effect is not new. \citet{Collados:2003} applied it to map the magnetic field azimuth of solar coronal filaments from forward-scattering polarization observations in the He {\sc i} 10830~\AA\, triplet. Moreover, \citet{Carlin:thesis} applied the same equation to the forward-scattering signals of the Ca {\sc ii}~8662~\AA\, line, which he calculated in a different 3D model of the solar atmosphere using the 1.5D radiative transfer approximation neglecting the horizontal components of the model's macroscopic velocities. In addition, he noted that the 8542~\AA\, line is also a suitable choice \citep{Carlin:2015}. We point out that the above-mentioned 1.5D approximation leads to the conclusion that the only way of producing forward scattering polarization is through the presence of inclined magnetic fields. However, in reality, the symmetry breaking produced by the horizontal inhomogeneities of the solar chromospheric plasma and by the spatial gradients in the horizontal components of its macroscopic velocity produce very significant forward-scattering signals in the \ion{Ca}{2}~8542~\AA\, chromospheric line without the need of inclined magnetic fields \citep[see][]{Stepan:2016}. The same applies to other chromospheric lines \citep[e.g.,][]{Jaume+2021}. 

The aim of the present paper is to investigate the reliability of Eq.~(\ref{chi_eq}) for determining the azimuth of the chromospheric magnetic field from the scattering polarization observed in the \ion{Ca}{2}~8542~\AA\, line, taking into account the impact of several telescope diameters and signal-to-noise ratios. To this end, we use the Stokes $Q$ and $U$ signals that result from full 3D radiative transfer calculations in the 3D model of \citet{Carlsson:2016}. 

It is important to emphasize the following points: 

(1) We take fully into account the effects of 3D non-LTE radiative transfer on the polarization of the emergent spectral line radiation, including the impact of the model's macroscopic velocity gradients. This is important because, as mentioned above, in full 3D radiative transfer the scattering polarization signals are strongly affected by the model's horizontal inhomogeneities and by the effects of spatial gradients in the three vectorial components of the plasma's macroscopic velocity \citep[e.g., the review by][]{2015IAUS..305..360S}. Neglecting these effects would not be suitable to study the quantitative impact of the instrumental effects on the inferred magnetic field azimuth, and this is precisely the main goal of this paper.

(2) The calculated polarization signals are caused by the combined action of scattering processes and the Hanle and Zeeman effects. 

(3) The azimuth of the model's magnetic field is not exactly constant along the LOS. 

(4) The strength of the model's magnetic field if not everywhere sufficiently strong so as to guarantee that the \ion{Ca}{2}~8542~\AA\, line is in the Hanle-effect saturation regime. 

(5) We take into account the impact of the instrumental degradation and noise, and in our simulated observations we consider several telescope apertures.

Although the solar chromosphere is more complex than any present 3D model, we think that the Stokes $Q/I$ and $U/I$ profiles calculated by \citet{Stepan:2016} in the 3D snapshot model of \citet{Carlsson:2016} include all the key physical ingredients needed to reach solid conclusions concerning the reliability of the basic formula of Eq. (\ref{chi_eq}) for mapping the azimuth of the magnetic field in quiet regions of the solar chromosphere.
      
\begin{figure*}
   \includegraphics[width=0.7\linewidth]{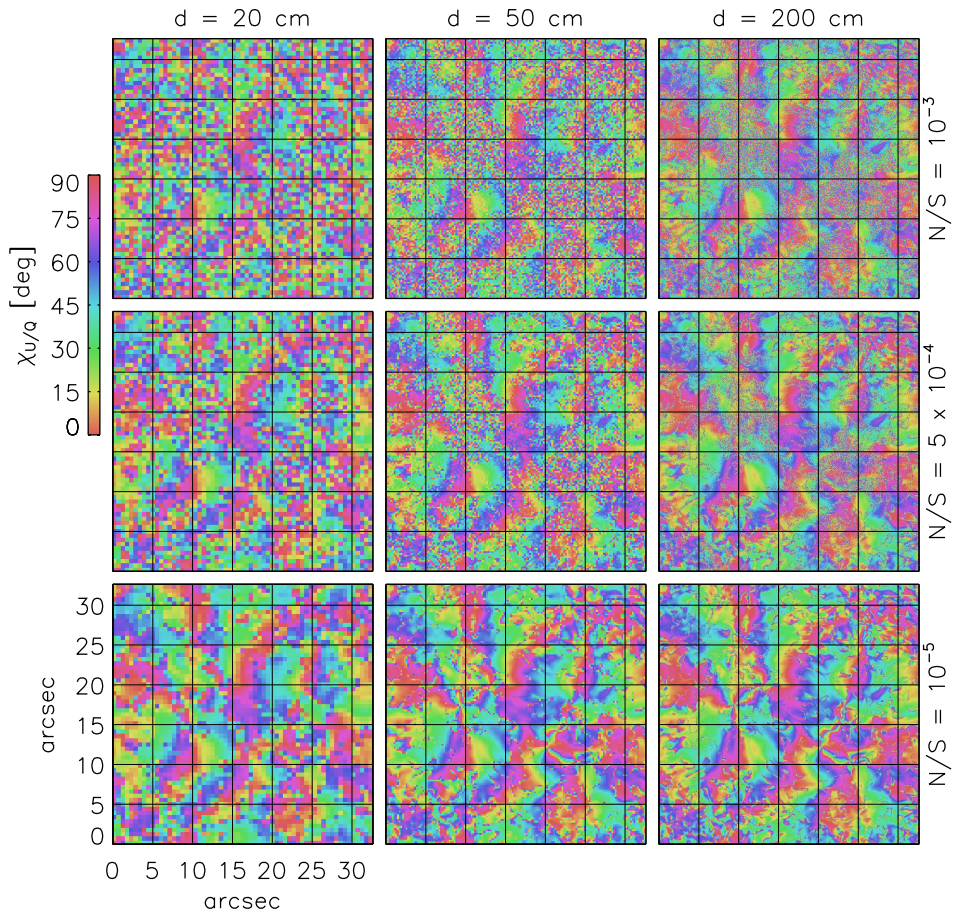}
   \caption{Maps of the magnetic field azimuth taking into account different telescope apertures (d) and 
   noise-to-signal (N/S) ratios. The pixel size used corresponds to the critical sampling of the theoretical spatial resolution.}
   \label{azimuths}%
\end{figure*}   
    
\begin{figure}
   \includegraphics[width=\linewidth]{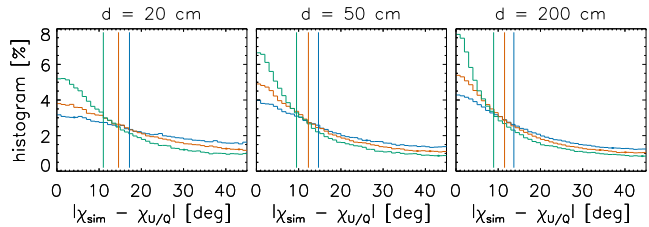}
   \caption{Histograms of the differences between the magnetic field azimuth $\chi_{U/Q}$ inferred 
   from the $U/Q$ ratio (see Eq.~(\ref{chi_eq})) and the actual azimuth $\chi_\mathrm{sim}$ of the model's 
   magnetic field at the atmospheric heights where the optical depth is unity at the center 
   of the \ion{Ca}{2}~8542~\AA\, line. Each plot corresponds to a different telescope diameter. Line colours indicate the N/S ratio: $10^{-3}$ (blue), $5 \times 10^{-4}$ (orange), and $10^{-5}$ (green). The thin vertical lines mark where we have a cumulative probability of 50\% for each telescope and N/S ratio.}
  \label{histograms}%
\end{figure}

\section{Analysis of the theoretical data}

We use a 3D snapshot model of the solar atmosphere resulting from a radiative-magnetohydrodynamics simulation of an enhanced network region \citep[see][]{Carlsson:2016}. We use snapshot 385 of the time-dependent simulation. The disk-center field of view covered by this 3D model is $32\farcs6 \times 32\farcs6$. The model has a grid size of approximately 48~km along the horizontal directions. 

The Stokes profiles of the emergent radiation in the \ion{Ca}{2}~8542~\AA\, line were calculated by \citet{Stepan:2016} using the 3D radiative transfer code PORTA\footnote{The public version of the PORTA radiative transfer code can be found at \url{https://gitlab.com/polmag/PORTA}.} \citep{Stepan:2013}, taking into account scattering processes and the Hanle and Zeeman effects. Figure~1 of \citet{Stepan:2016} and Fig.~4 in \citet{Jurcak:2018a} provide information on the physical conditions of the 3D model atmosphere at the corrugated surface where the optical depth is unity at the center of the \ion{Ca}{2}~8542~\AA\, line, for the disk center line of sight.

The two leftmost panels of Fig.~\ref{fig1} show the amplitudes of the calculated $Q/I$ and $U/I$ profiles. The middle panel gives the map of the magnetic field azimuth obtained using Equation~(\ref{chi_eq}), after restricting the resulting $\chi_{U/Q}$ values between $0^\circ$ and $90^\circ$ due to the azimuth ambiguities that follow from Eq.~(\ref{chi_eq})\footnote{For a detailed discussion of the magnetic field ambiguities resulting from the inference of the polarization observed in spectral lines we refer the reader to sections 1.9 and 11.7 of \cite{Egidio:2004}}. The next panels show the actual values of the model's magnetic field azimuth ($\chi_\mathrm{sim}$) at the corrugated surface of line-center optical depth unity in the vertical direction and a probability density function (PDF) that quantifies the accuracy of the inferred magnetic field azimuth. 

In order to evaluate the reliability of the inferred magnetic field azimuth, $\chi_{U/Q}$, we sum the probabilities within the red dashed lines in the PDF panel of Fig.~\ref{fig1}; i.e., we calculate the total probability that $|\chi_{U/Q} - \chi_\mathrm{sim}| < 10^\circ$. To investigate the reliability of $\chi_{U/Q}$ depending on different observing setups, we degrade the theoretical data considering various N/S ratios and telescope apertures. To this end, we: (1) convolve the calculated Stokes profiles with the spatial PSF of an ideal telescope having an unobscured entrance pupil ranging from 10~cm to 400~cm in diameter; (2) re-sample the resulting maps of Stokes profiles to pixel sizes corresponding to half of the theoretical spatial resolution of the assumed telescope diameter; (3) add white noise to all the Stokes profiles using a Gaussian distribution with $\sigma$ ranging from $5 \times 10^{-3}$ to $10^{-5}$.

The $Q$ and $U$ signals used in Equation~\ref{chi_eq} result from averaging those corresponding to the nine wavelength points located around the minimum of the Stokes $I$ profile. This method is applicable also when the  individual $Q$ and $U$ signals are dominated by noise. We did not modify the 12~m\AA\, wavelength sampling of the theoretical Stokes profiles, which corresponds to a spectral resolution $R \sim 350\,000$. Averaging nine wavelength points effectively decreases the noise level by a factor three. This might not be an option if the spectrograph used has lower spectral resolution. In such a case, such low N/S ratios would be achieved with the same exposure time because the spectrograph automatically integrates the Stokes signals in wavelength. Note that the spatial grid size of the 3D model atmosphere is comparable to the diffraction limit of a telescope with an entrance pupil of 160~cm, and that such spatial resolution is significantly worse than the diffraction limit of the DKIST and EST telescopes.

\section{Reliability of the azimuth determination}

Figure~\ref{azimuths} shows the resulting maps of the inferred $\chi_{U/Q}$ magnetic field azimuth for different telescope diameters (columns) and N/S ratios (rows). There is a clear trend of improvement with increasing telescope aperture and decreasing N/S ratio. The lower right $\chi_{U/Q}$ map is nearly identical to the map of the model's $\chi_\mathrm{sim}$ azimuth shown in Fig.~\ref{fig1}. 

\begin{figure*}
   \includegraphics[width=0.7\linewidth]{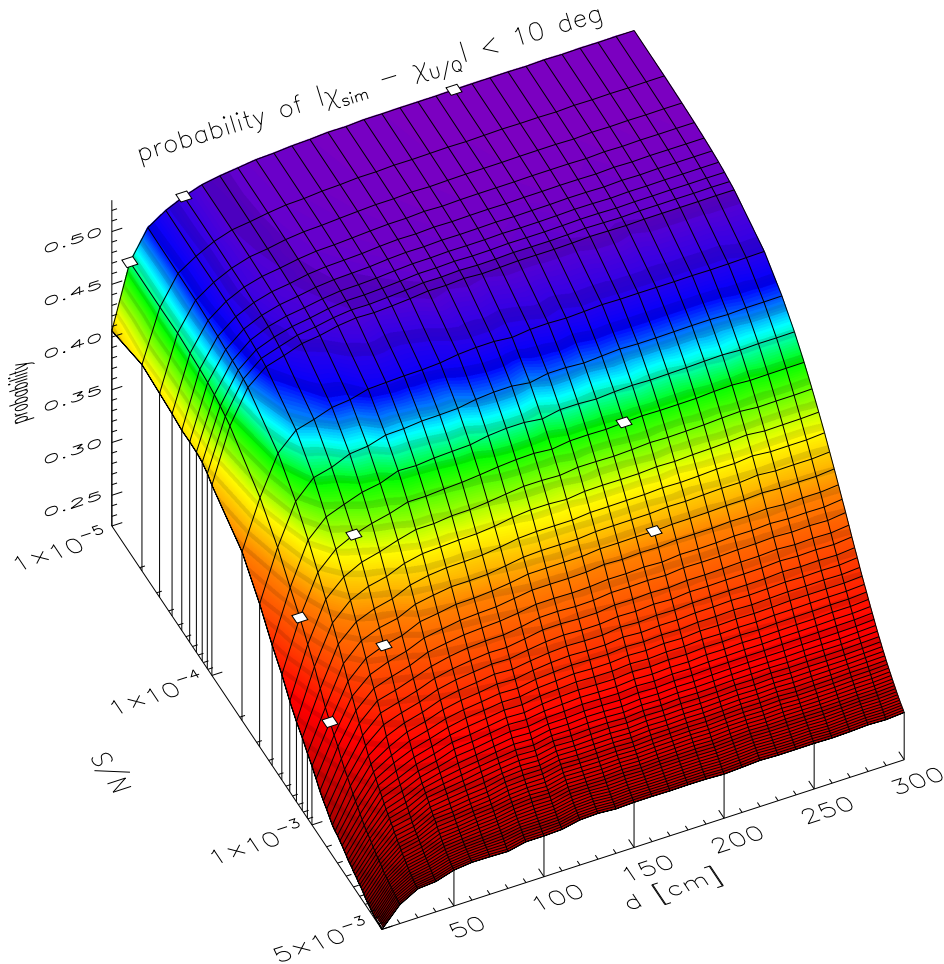}
   \caption{The probability of $|\chi_{U/Q} - \chi_\mathrm{sim}| < 10^\circ$ for different telescope diameters and N/S levels. The white points indicate the cases considered in Figs.~\ref{azimuths} and~\ref{histograms}. The colored isosurface is interpolated between the grid points that correspond to the black-curves crossings.}
  \label{surface}%
\end{figure*}

In order to compare the inferred $\chi_{U/Q}$ maps with the $\chi_\mathrm{sim}$ ones we show in Fig.~\ref{histograms} the histograms of the $|\chi_{U/Q} - \chi_\mathrm{sim}|$ differences for each of the maps shown in Fig.~\ref{azimuths}. Note that a random distribution of $\chi_{U/Q}$  would produce a straight line at 2.22\% in Fig.~\ref{histograms}. Figure~\ref{surface} shows the cumulative probability of $|\chi_{U/Q} - \chi_\mathrm{sim}| < 10^\circ$ for a range of N/S ratios and telescopes with diameters $d \leq 300$~cm.

The different columns of Fig.~\ref{azimuths} illustrate the importance of the telescope diameter. With $d = 50$~cm we already have a spatial resolution that captures reasonably well the global structure of the magnetic field (the pixel size is roughly three times larger that the model's grid size). With increasing telescope aperture, we do not gain a significant improvement in the reliability of $\chi_{U/Q}$. The data with the lowest N/S ratio of $10^{-5}$ shows the highest probability of $|\chi_{U/Q} - \chi_\mathrm{sim}| < 10^\circ$ for telescope diameters larger than $\sim 150$~cm, for which the spatial resolution is comparable to the model's grid size (Fig.~\ref{surface}). For such telescopes the cumulative probability is around 53\%, decreasing only to 51\% and 42\% for telescope diameters of 50~cm and 10~cm, respectively. 

Clearly, the N/S ratio has an important impact on the reliability of the inferred magnetic field azimuth   (see Figs~\ref{azimuths} and \ref{surface}). For telescopes with $d = 200$~cm the cumulative probability value drops from 53\% for a N/S ratio of $10^{-5}$ to 45\% and 39\% for N/S values of $5 \times 10^{-4}$ and $10^{-3}$, respectively. For larger N/S ratios we approach a cumulative probability value of 23\%, which is equivalent to a random distribution of the $\chi_{U/Q}$ azimuths. These results indicate that to some extent we can sacrifice the spatial resolution in order to decrease the noise level in our simulation of measured $Q$ and $U$ signals. 

\section{Summary and conclusions }

We have investigated the reliability of a simple formula for obtaining maps of the magnetic field azimuth from the linear polarization signals of the \ion{Ca}{2}~8542~\AA\, line. To this end, we use the Stokes profiles calculated with the PORTA radiative transfer code in a 3D snapshot model of the quiet solar chromosphere, taking into account scattering processes and the Hanle and Zeeman effects, as well as the Doppler shifts produced by the model's macroscopic velocities. The method is based on the well-known Eq.~(\ref{chi_eq}), which can in principle be applied to spectral lines that are in the saturation regime of the Hanle effect, as it approximately happens with the \ion{Ca}{2}~8542~\AA\, line in the solar chromosphere. Our emergent Stokes profiles result from full 3D radiative transfer calculations, which is the reason why in our case an inclined magnetic field is not the only cause of symmetry breaking (i.e., we do not use the 1.5D approximation). In addition, we have  accounted for the limited spatial resolution corresponding to various telescope diameters and the impact of several signal-to-noise ratios on the simulated observations.
  
In the 3D snapshot model atmosphere we have used, the magnetic field azimuth is not strictly constant along the LOS and the magnetic field is not everywhere sufficiently strong so as to guarantee that the \ion{Ca}{2}~8542~\AA\, line is always in the saturation regime of the Hanle effect. The real solar chromosphere is more complex than any present 3D model, but our theoretical $Q/I$ and $U/I$ signals are sufficiently realistic so as to argue that we can use them to reach reasonable conclusions regarding the reliability of the applied method for inferring the magnetic field azimuth from real observations. Given that 3D radiative transfer calculations are computationally costly, in this investigation we have used a single 3D snapshot model (i.e., we have not accounted for the impact of the temporal evolution of the solar chromospheric plasma). Nevertheless, with the new generation of large-aperture solar telescopes and spectropolarimeters the exposure time needed to detect the \ion{Ca}{2}~8542~\AA\, polarization is expected to be around one minute (see below).     

To quantify the reliability of the inferred magnetic field azimuths ($\chi_{U/Q}$) we use the cumulative probability that the obtained $\chi_{U/Q}$ values are within $10^\circ$ of the model's magnetic field azimuth at the heights where the line-center optical depth is unity in the \ion{Ca}{2}~8542~\AA\, line ($\chi_\mathrm{sim}$). For the sake of simplicity, we have presented our analysis for the forward-scattering geometry of the disk-centre line of sight.

Our results show that the azimuth estimation is very sensitive to the noise level of the spectropolarimetric observation and that, to some extent, we can sacrifice the spatial resolution to improve the polarimetric sensitivity. However, this conclusion has been reached using Stokes profiles from radiative transfer calculations in a 3D model atmosphere having a relatively large grid size, which is comparable to the spatial resolution of telescopes with a diameter $\sim 150$~cm. Moreover, the simulation box contains a relatively simple magnetic field configuration, while the magnetic structure of the real solar chromosphere is probably more complex. A recent investigation contrasting the calculated Stokes profiles \citep{Stepan:2016} with disk-center spectropolarimetric observations of the \ion{Ca}{2}~8542~\AA\, line indicates that the structuring of the solar chromospheric plasma is significantly more complicated than in the used 3D model \citep{Jurcak:2018a}. New spectropolarimetric observations in the \ion{Ca}{2}~8542~\AA\, line at all positions on the solar disk further underscore the scientific interest of its scattering polarization signals \citep{Harvey:2020}. Clearly, we would certainly benefit from new-generation solar telescopes like DKIST and EST, which will hopefully provide us with spectropolarimetric data of high spatial resolution and polarimetric sensitivity. 

A key advantage of large-aperture solar telescopes, like DKIST and EST, is their collecting power. For a fixed resolution element, the photon flux scales with the area of the primary mirror. Therefore, with any of such telescopes ($d=4$~m) the exposure time needed to achieve a given N/S ratio is a factor 16 lower than with a 1~m telescope. Assume we want to achieve a N/S ratio of $10^{-4}$ at the core of the \ion{Ca}{2}~8542~\AA\, line with a spectral resolution $R = 40\,000$ (which roughly corresponds to the spectral resolution of averaging nine wavelength points, as in our study), and a spatial sampling of 0\farcs1. With the Visible Spectro-Polarimeter (ViSP, one of the DKIST first-generation instruments) the necessary exposure time would be around one minute, according to the ViSP Instrument Performance Calculator. However, with a 1.5~m telescope the exposure time would be 7 times longer. This simple example highlights further the need of large aperture solar telescopes for making feasible precise spectropolarimetric observations of the highly dynamic solar chromosphere. In addition, it is very important to develop integral field unit instruments capable of achieving what the slit-based ViSP instrument allows, but simultaneously over a sufficiently large two-dimensional field of view.

In this investigation we have not taken into account possible spatial correlations between the nearby pixels, nor the fact that the azimuth ambiguity could be resolved by using constraints from other methods. Such extension of our work is left for a future investigation.

\acknowledgments

J.J. and J.\v{S}. acknowledge financial support by the Grant Agency of the Czech Republic through grant 19-20632S and project RVO:67985815. J.T.B. acknowledges the funding received from the European Research Council (ERC) under the European Union’s Horizon 2020 research and innovation programme (Advanced Grant agreement No. 742265), as well as through the projects PGC2018-095832-B-I00 and PGC2018-102108-B-I00 of the Spanish Ministry of Science, Innovation and Universities. Likewise, J.T.B. and J.\v{S}. acknowledge the support received from the Swiss National Science Foundation through grant CRSII5-180238. The 3D radiative transfer calculations were carried out with the MareNostrum supercomputer of the Barcelona Supercomputing Centre (National Supercomputing Centre, Barcelona, Spain).

\bibliography{manuscript}{}
\bibliographystyle{aasjournal}

\end{document}